\documentclass[prb,reprint,showpacs,preprintnumbers,amsmath,amssymb]{revtex4-1}
\usepackage{graphicx}
\usepackage{dcolumn}
\usepackage{bm}
\usepackage[usenames]{color}
\usepackage{amssymb}
\usepackage{amsfonts}
\usepackage{ae, aeguill}
\usepackage{multirow}
\usepackage{color}
\usepackage{booktabs}
\newcommand{\head}[1]{\textnormal{\textbf{#1}}}

\begin{document}

\title{Parimagnetism in RCo$_2$ series (R=Dy, Ho, and Tm)}

\author{C. M. Bonilla}
\email{mbonilla@unizar.es} \affiliation{Instituto de Ciencia de
Materiales de Arag\'{o}n and Departamento de F\'{i}sica de la
Materia Condensada, CSIC - Universidad de Zaragoza,
 Pedro Cerbuna
12, E-50009 Zaragoza, Spain}
\author{J. Herrero-Albillos}
\affiliation{Fundaci\'{o}n ARAID, Paseo Mar\'{i}a Agust\'{i}n 36, E-50004 Zaragoza, Spain} \affiliation{Centro Universitario de la Defensa, Ctra. de Huesca s/n, E-50090 Zaragoza, Spain}\affiliation{Instituto de Ciencia de Materiales de Arag\'{o}n, CSIC - Universidad de Zaragoza, Pedro Cerbuna 12, E-50009 Zaragoza, Spain }
\author{A. I. Figueroa}
\author{C. Cast\'{a}n-Guerrero}
\author{J. Bartolom\'{e}}
\affiliation{Instituto de
Ciencia de Materiales de Arag\'{o}n and Departamento de F\'{i}sica de la
Materia Condensada, CSIC - Universidad de Zaragoza,  Pedro Cerbuna
12, 50009 Zaragoza, Spain}
\author{I. Calvo-Almaz\'{a}n}
\affiliation{Institute Laue Langevin, 38042, Grenoble, France}
\affiliation{Instituto de
Ciencia de Materiales de Arag\'{o}n and Departamento de F\'{i}sica de la
Materia Condensada, CSIC - Universidad de Zaragoza,  Pedro Cerbuna
12, 50009 Zaragoza, Spain}
\author{D. Schmitz}
\author{E. Weschke}
\affiliation{Helmholtz-Zentrum Berlin f$\ddot{u}$r Materialien und
Energie GmbH, Albert-Einstein-Str. 15, 12489 Berlin, Germany}
\author{L. M. Garc\'{i}a}
\author{F. Bartolom\'{e}}
\affiliation{Instituto de
Ciencia de Materiales de Arag\'{o}n and Departamento de F\'{i}sica de la
Materia Condensada, CSIC-Universidad de Zaragoza,  Pedro Cerbuna
12, 50009 Zaragoza, Spain}

\date{\today}

\begin{abstract}

X-ray circular magnetic dichroism (XMCD), longitudinal ($\chi_{ac}$) and transverse (TS) ac magnetic susceptibility have been measured in several members of the $R$Co$_2$ series ($R$ = Dy, Ho, and Tm) as a function of temperature and applied magnetic field. We show that parimagnetism is a general behavior along the $R$Co$_2$ ferrimagnetic series ($R$ being a heavy rare earth ion). XMCD results evidence the presence of two compensation temperatures, defining two different parimagnetic configurations, which is a fully unexpected result. The inverse $\chi_{ac}$ curve exhibits a deviation from Curie-Weiss behavior which is recovered under applied magnetic field. The large excess of polarizability above the critical temperature proves the existence of an enhanced effective moment due to the presence of short range magnetic correlations, which are also observed in TS measurements. The combination of TS and XMCD measurements allows to depict new magnetic phase diagrams for the $R$Co$_2$ series. A new scenario allowing to understand the observed phenomenology as a Griffiths phase-like behavior is proposed, where the amorphous $R$Co$_2$ represents the undiluted system case.

\end{abstract}


\maketitle


\section{\label{Intro}INTRODUCTION}

Among the newly discovered phenomena related to short-range magnetic correlations, an interesting example is the parimagnetic phase observed in ErCo$_2$. \cite{Herrero-Albillos2007PRB76} Parimagnetism consists in the antiparallel alignment of the net magnetization of each sublattice in the magnetically disordered, high temperature phase $(T>T_c)$, of a two-sublattice system under an applied magnetic field, differently to the usual paramagnetic phase where the two sublattices would present net magnetizations parallel to the applied magnetic field. This phenomenon indeed contradicts the usual interpretation of the $R$Co$_2$ series as simple exchange-enhanced paramagnets with a rare-earth dominated magnetism.\cite{StewartPCSSP1984, BlochPRB1970, DeportesPSSB1974}

The hierarchy of exchange magnetic interactions in $R$Co$_2$ is $J_{\mathrm{Co-Co}}\gg J_{\mathrm{R-Co}}$,\cite{RhyneJMMM1987,CastetsJMMM1980} while $J_{\mathrm{R-R}}$ is usually neglected. Therefore, the formation of Co short-range magnetic correlations in the paramagnetic phase is energetically favored.  Indeed, Co magnetic correlations have been previously identified at the origin of some anomalies in the transport properties of $R$Co$_2$, but it had been ascribed to critical fluctuations,\cite{Gratz_JPCM2001} near the Curie temperature. Nevertheless, previous works on ErCo$_2$ and HoCo$_2$; including small angle neutron scattering (SANS),\cite{Herrero-Albillos2007PRB76, Bonilla_2012} longitudinal ($\chi_{\mathrm{ac}}$)\cite{Herrero-Albillos_JPCM2009} and transverse (TS)\cite{Bonilla_2012, Figueroa_JAP2011} magnetic susceptibilities and muon spin relaxation spectroscopy ($\mu$SR),\cite{Bonilla_PRB2011} among others, have shown the presence of magnetic correlations not only near $T_c$, but also at much higher temperatures,. Magnetic correlations at temperatures above the ferrimagnetic ordering in ErCo$_2$ had also been identified in the neutron diffraction pattern\cite{Pirogov_APA2002,Podlesnyak_PRB2002} and in electrical resistivity measurements.\cite{Garcia_JMMM_2001} Magnetization and transport properties show the fingerprint of high temperature magnetic correlations in related compounds, such as Er(Co$_x$Ni$_{1-x}$)$_2$\cite{Soares_JMMM1999} and Er(Co$_x$Ti$_{1-x}$)$_2$.\cite{Oner_JAC2008} Other intermetallic Laves phases do also show magnetic correlations at high temperatures compared with $T_C$: D\'eportes and coworkers\cite{DeportesJAP1981} claimed the existence of short-range magnetic order at four times $T_C$ in CeFe$_2$, evidencing the importance of transition metal interactions in the paramagnetic phase of these intermetallic compounds.

Previous work in ErCo$_2$ allowed to identify a Griffiths - like phase within its magnetically disordered phase.\cite{Herrero-Albillos_JPCM2009} A Griffiths phase can be seen as the realization of short-range order, within a diluted magnet of a given concentration $x$, lying between the percolation concentration, $x_{\mathrm p}$, and the undiluted ``pure'' system, $x=1$.\cite{Griffiths_PRL_69} The Griffiths phase appears below a given temperature, $T_{\mathrm G}$, above the critical ordering temperature of the diluted system and that of the pure one: $T_c^x<T_{\mathrm G}<T_c^1$.  The short range order is a kind of low-temperature remnant of the magnetic order of the undiluted system, in the temperature range at which the system would order spontaneously if it were not diluted. This is the scenario in most of the physical realizations of Griffiths phases claimed up to now, as in germanates  \cite{Ouyang_PRB_2006,Magen_PRL_2006,Pereira_PRB_2010,Perez_PRB_2011}, manganites\cite{Salamon_PRB_2003,Salamon_PRL_2002,JiangPRB2007} and alloys.\cite{Herrero-Albillos_JPCM2009,Castro_PRL_1998,Oner_JMMM2012,Perez_PRB_2011} The ac-susceptibility curve obtained for ErCo$_2$ shows the usual deviation from a Curie Weiss behavior below a certain temperature  which is, however, recovered under applied magnetic field.\cite{Salamon_PRL_2002,Ouyang_PRB_2006,Magen_PRL_2006,Perez_PRB_2011} Such extreme sensitivity to the applied field is characteristic of a Griffiths phase.\cite{Salamon_PRB_2003,JiangPRB2007} However, the physical origin of a Griffiths - like phase in a pure compound like ErCo$_2$ is not easy to understand.

From previous experimental work \cite{Herrero-Albillos_JPCM2009,Herrero-Albillos2007PRB76,Bonilla_PRB2011,Figueroa_JAP2011,Bonilla_2012,Herrero-Albillos_JMMM2007,Fernando_JMMM2004,Herrero-Albillos_JMMM_Cuello2007,Misek_JAP_2012}
parimagnetism in ErCo$_2$ is a consequence of short-range correlations between Co magnetic moments and it is natural to investigate whether this phenomenon is present among other ferrimagnetic $R$Co$_2$.

A systematic X-ray magnetic circular dichroism (XMCD) study together with magnetic susceptibility measurements performed in $R$Co$_2$ with $R$= Dy, Ho and Tm, are presented in order to obtain a more complete description of the magnetism above $T_c$ in these compounds. Our work indicates that parimagnetism observed in the ErCo$_2$ system is found in the $R$Co$_2$ $R$= Dy, Ho and Tm, members of the series too, and the effect of short-range correlations are even stronger than those found in ErCo$_2$, showing some unexpected features reminiscent of compensation points in rare-earth transition metal compounds.

\begin{figure*}[!htb]
\begin{center}
\includegraphics[width=1\textwidth]{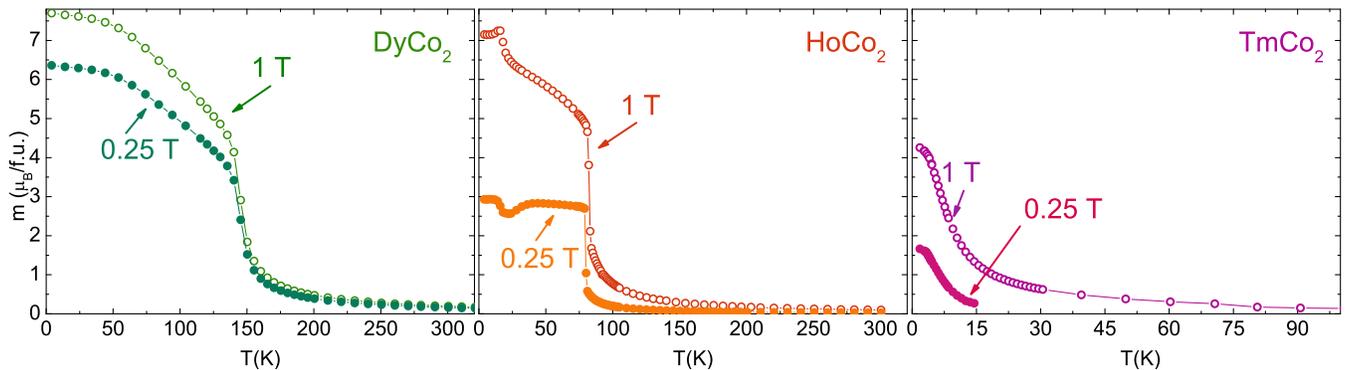}
\caption{\label{fig:M(T)RCo2}(Color online) Magnetization as
function of temperature at $\mu_0H$=0.25 T (full symbol) and $\mu_0H$=1 T (open
symbol). From left to right for ferrimagnetic systems DyCo$_2$,
HoCo$_2$ and TmCo$_2$.}
\end{center}
\end{figure*}

\section{\label{Samples}SAMPLE CHARACTERIZATION AND EXPERIMENTAL DETAILS}

Polycrystalline ingots of $R$Co$_2$ with $R$= Dy, Ho and Tm were
prepared following a standard procedure by melting together
the metallic precursors in an arc furnace under Ar atmosphere. An excess of 1$\%$ of rare earth has been added to obtain a final sample with the correct stoichiometry, taking into account the amount of rare earth which is evaporated during the melting process. The samples were annealed under Ar atmosphere at $850$
$^{\circ}$C for twelve days in order to improve the homogeneity.
X-ray diffraction analysis on powdered samples was performed in a
Rigaku RTO 500RC diffractometer with Bragg-Brentano geometry using K$\alpha$ Cu radiation at room temperature. The Rietveld
analysis shows very well crystallized samples, as shown in previous works\cite{Herrero-Albillos2007PRB76} and no impurities within the 2$\%$ of accuracy of
powder diffraction methods.

A complete magnetic characterization has been performed on the powdered samples. The longitudinal ac susceptibility has been measured under applied fields up to $\mu_0H$=5 T in a SQUID Quantum Design magnetometer from $T$=4 K to $T$=400 K. Magnetization ($M$) (Fig.~\ref{fig:M(T)RCo2}) and ac
susceptibility $\chi_{ac}$  as function of temperature measured for the three ferrimagnetic systems show critical temperatures $T_c$= 136 K, 78 K and 4.6 K for DyCo$_2$, HoCo$_2$, and TmCo$_2$, respectively, which coincide with those reported in the literature.\cite{Khmelevskyi_JPCM_2000}

The presence of short-range magnetic correlations above $T_c$ has been studied by means of transverse susceptibility (TS). These measurements were performed using a radiofrequency (RF) self-resonant circuit oscillator using CMOS transistors coupled with a LC tank, operating at a frequency of about 12 MHz.\cite{FigueroaJMMM2012} The sample and coil set are designed to fit into a commercial Quantum Design PPMS probe such that the perturbing RF magnetic field inside the coil ($\approx$ 10$^{-3}$ T) is oriented perpendicular to the external dc magnetic field $H_{dc}$ supplied by the PPMS superconducting magnet, thus setting the transverse geometry. In a TS measurement at a fixed temperature, the resonant frequency $f$ is recorded as the $H_{dc}$ field is swept from positive to negative saturation (unipolar scan) and then back to positive (bipolar scan). Since the change in frequency of the circuit, $\Delta f$, is a direct consequence of the change in inductance as the sample is magnetized, $\Delta f$ is proportional to the variation of the transverse susceptibility of the sample, $\Delta \chi_{T}$, the magnitude of physical interest. In the present study, the quantity $[\Delta \chi_{_T}/\chi_{_T}]\%=[\chi_{_T}(H_{dc})-\chi_{_T}^{sat}]/\chi_{_T}^{sat}\times100\%$ as function of the applied field $H_{dc}$ is the one considered, where $\chi_{_T}^{Sat}$ is the transverse susceptibility at the saturating field $\mu_0H^{sat}$=1 T, within a temperature range 2.5 K$<T<$ 300 K. Details on the technique and the circuit can be found in Ref. \onlinecite{FigueroaJMMM2012}.

The skin depth, $\delta$, of the $R$Co$_2$ compounds has been calculated by the well-known formula $\delta = \sqrt{\rho \, / \,2 \pi \!f \!\mu}$ where $\rho$ is the resistivity, $\mu$ the magnetic permeability and $f$ the excitation frequency. The values range between 30 $\mu$m at low temperatures and 180 $\mu$m for $T>T_c$ up to room temperature. In order to assure that the RF magnetic field penetrates the entire grain the sample has been powdered to reach a grain size of average diameter $\sim 1.5 \; \mu$m, well below $\delta$, to avoid the formation of eddy currents on the surface.

X-ray absorption spectroscopy (XAS) and X-ray magnetic circular dichroism (XMCD) measurements were carried out at the UE46-PGM1 beamline, at BESSY synchrotron facility, Berlin. The measurements were performed at the Co L$_{2,3}$ and the Dy, Ho and Tm M$_{4,5}$ absorption edges with a polarization rate of 0.9. The temperature was varied between 5 K to 350 K under applied fields between $\mu_0H$=0.25 T and $\mu_0H$=6 T. The detection method used was total electron yield. In order to prevent the spurious signals due to surface oxidation, the polycrystalline ingots were cleaved in an ultra high vacuum chamber just before starting its exposure to the X-ray beam.

\section{\label{Results}EXPERIMENTAL RESULTS}

\subsection{\label{XMCD} X-ray magnetic circular dichroism}

\begin{figure}[!htb]
\begin{center}
\includegraphics[width=0.47\textwidth]{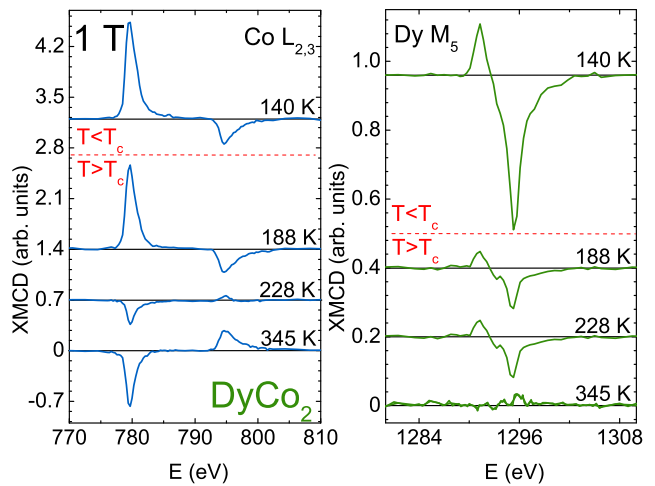}
\caption{\label{fig:XMCDDyCo2}(Color online) XMCD spectra at the
Co L$_{2,3}$ (left) and Dy M$_{5}$ (right) edges in DyCo$_2$ for
selected temperatures at $\mu_0H$=1 T. Dashed line separates the XMCD
spectra measured above and below $T_c$.}
\end{center}
\end{figure}
%

\begin{figure}[!htb]
\begin{center}
\includegraphics[width=0.47\textwidth]{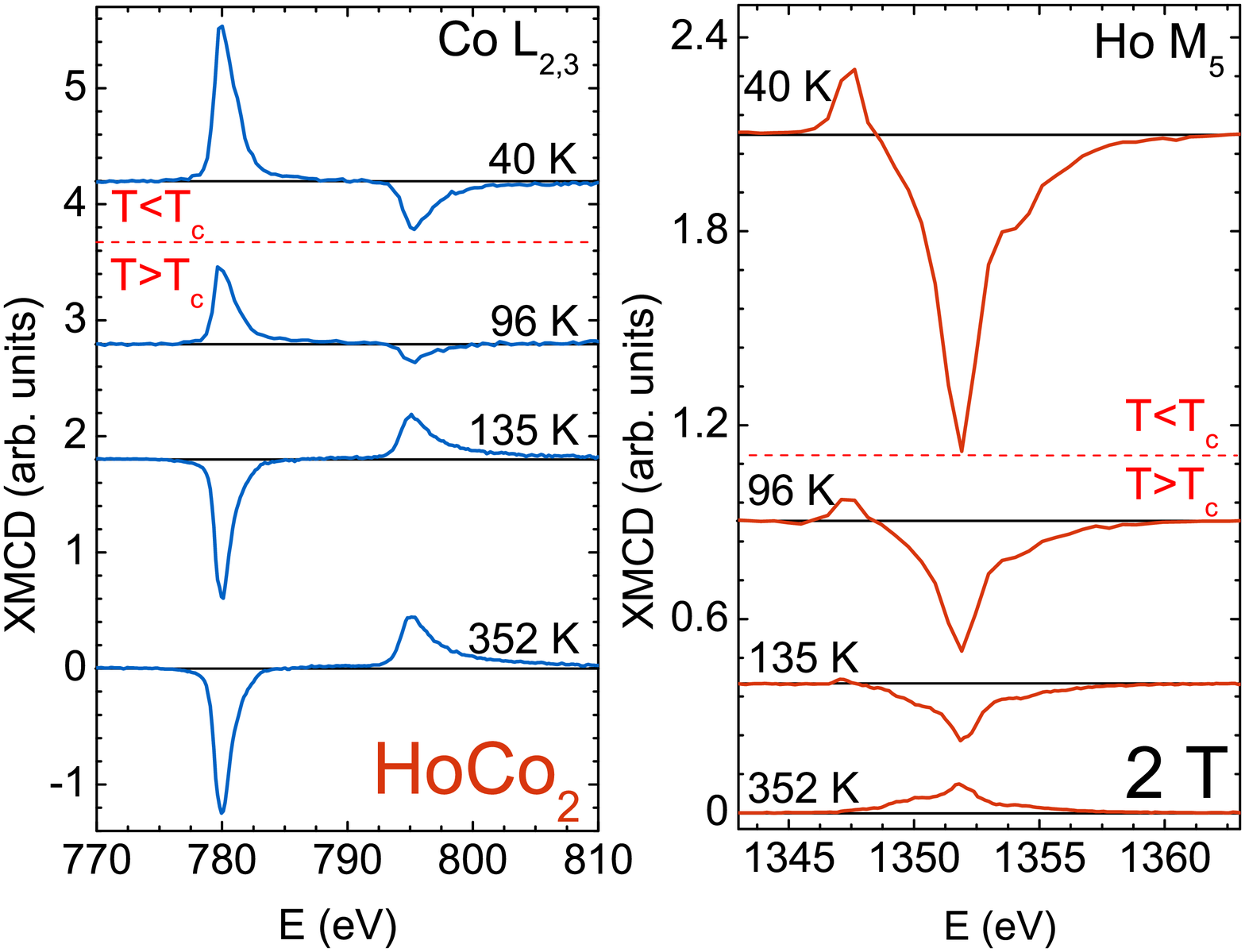}
\caption{\label{fig:XMCDHoCo2}(Color online) XMCD spectra at the
Co L$_{2,3}$ (left) and Ho M$_{5}$ (right) edges in HoCo$_2$ for
selected temperatures at $\mu_0H$=2 T. Dashed line separates the XMCD
spectra measured above and below $T_c$.}
\end{center}
\end{figure}
%

\begin{figure}[!htb]
\begin{center}
\includegraphics[width=0.47\textwidth]{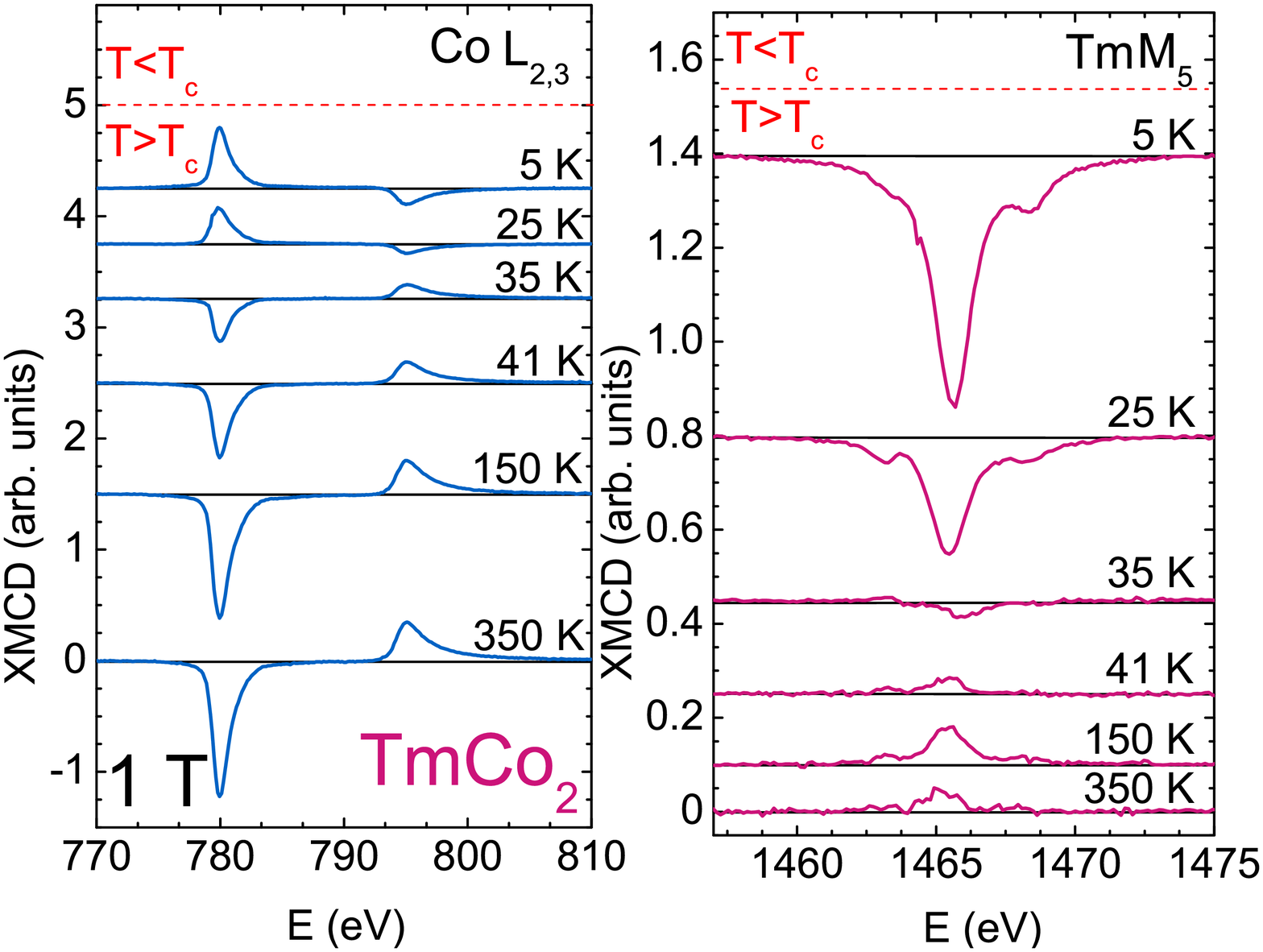}
\caption{\label{fig:XMCDTmCo2}(Color online) XMCD spectra at the
Co L$_{2,3}$ (left) and Tm M$_{5}$ (right) edges in TmCo$_2$ for
selected temperatures at $\mu_0H$=1 T. Dashed line separates the
spectra measured above and below the critical temperature.}
\end{center}
\end{figure}
%

\begin{figure}[!htb]
\begin{center}
\includegraphics[width=0.47\textwidth]{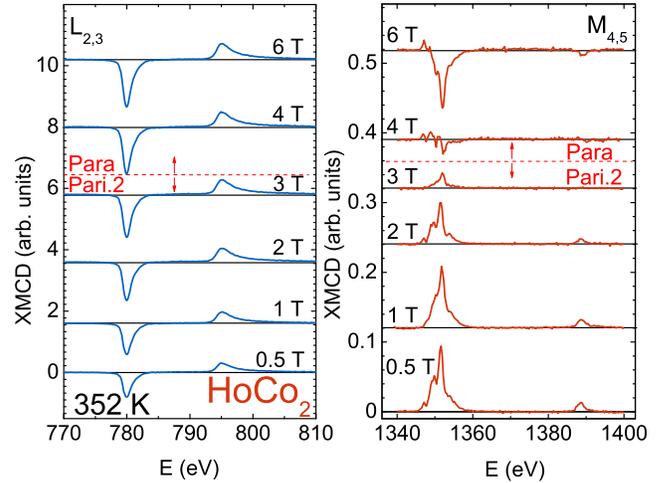}
\caption{\label{fig:IsofieldHoCo2}(Color online) XMCD spectra at
the Co L$_{2,3}$ (left) and Ho M$_{4,5}$ (right) edges in HoCo$_2$
for selected fields at $T$=352 K. Dashed line separates the
paramagnetic and parimagnetic-2 configurations.}
\end{center}
\end{figure}

DyCo$_2$, HoCo$_2$ and TmCo$_2$ have been studied by means of XMCD spectroscopy to obtain the element-specific magnetization in these materials. The XMCD signals show the usual spectral shape, for Co L$_{2,3}$ and $R$ M$_{4,5}$ edges, as is mentioned in Refs. \onlinecite{Chen_PRL} and \onlinecite{Goedkoop_thesis1989} respectively. The XMCD signals at the $R$ M$_4$ edges have in every case the same sign and (as expected) much reduced intensity with respect to the the $R$ M$_5$ edge ones, and will be, therefore, omitted in the majority of this work, for the sake of clarity.  Figs.~\ref{fig:XMCDDyCo2}, \ref{fig:XMCDHoCo2} and \ref{fig:XMCDTmCo2} show XMCD spectra for Co L$_{2,3}$ edges (left) and Dy, Ho and Tm M$_{5}$ edges (right), respectively, at selected temperatures and at an applied magnetic field of $\mu_0H$=1 T for DyCo$_2$ and TmCo$_2$ and $\mu_0H$=2 T for HoCo$_2$.

The Co absorption at the L$_{2,3}$ edges has been normalized using the usual 2:1 branching ratio.\cite{Thole_EPL_1987} The normalization of the M$_{4,5}$, has been performed by setting to 1 the maximum of the unpolarized absorption (obtained by averaging right and left polarized light XAS). The net magnetic moment of the rare-earth sublattice has been assumed to be proportional to the area under the XMCD signal at $R$ M$_5$ edge,\cite{Pizzini_JES1997} in such a way that negative values for the integral calculated from the XMCD spectrum correspond to a positive sign of the $R$ net magnetic moment. On the other hand, given $A$ and $B$, the areas under the Co XMCD curves at the L$_3$ and L$_2$ edges respectively, the magnetic moment per Co atom is proportional to $-5 A + 4 B$.\cite{Chen_PRL} The spectra at the top of each panel in Fig.~\ref{fig:XMCDDyCo2} and \ref{fig:XMCDHoCo2}, show the XMCD signals recorded below $T_c$ in DyCo$_2$ and HoCo$_2$, respectively. The sign of these spectra demonstrates the ferrimagnetic ordering of these phases, with the larger rare-earth net magnetic moment parallel to the applied field and the Co one, much smaller, antiparallel to it. XMCD could not be measured below $T_c$ on TmCo$_2$, as $T_c$ lies below the lowest temperature achievable at the experimental setup.

\begin{figure}[!htb]
\begin{center}
\includegraphics[width=0.48\textwidth]{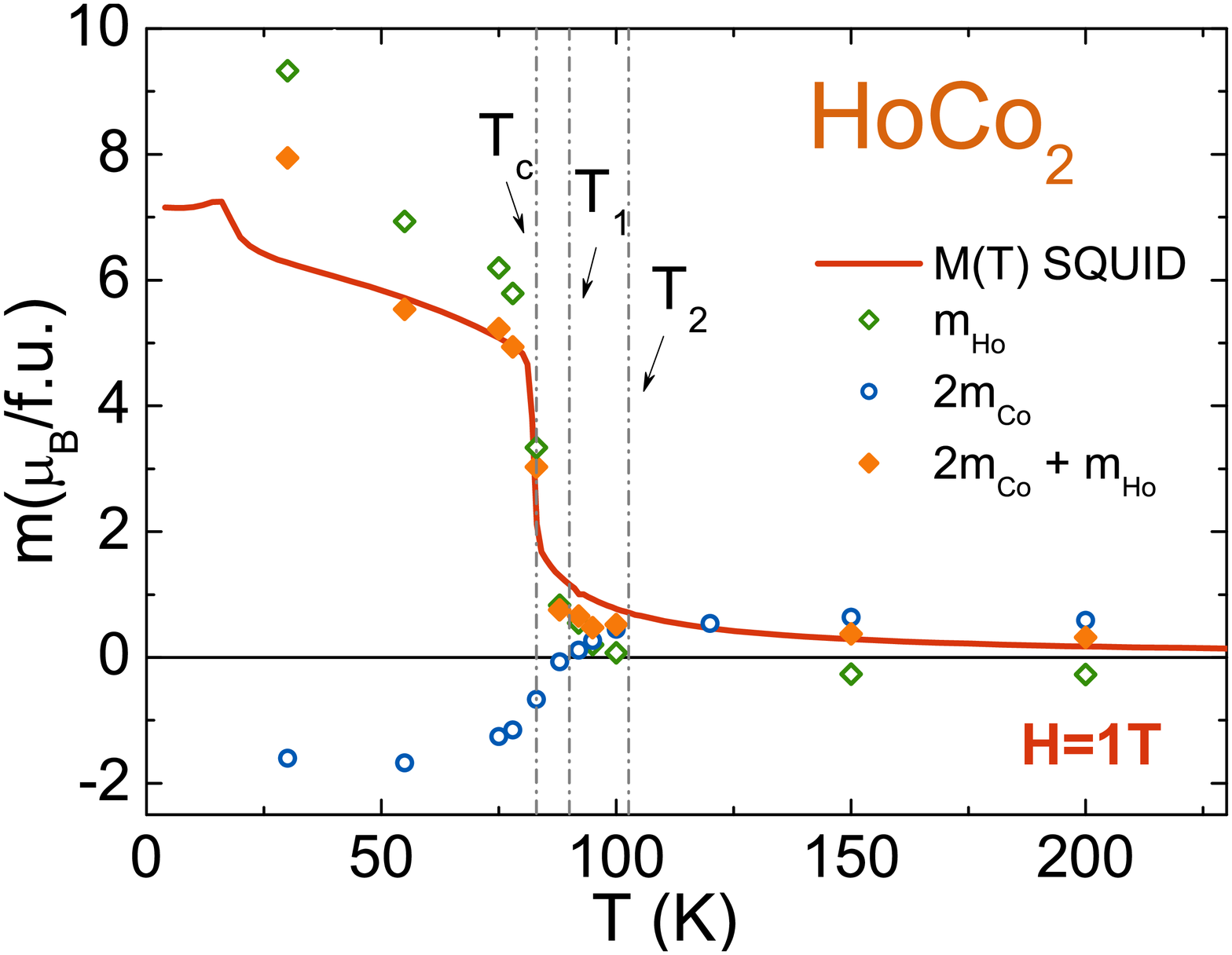}
\caption{\label{fig:M(T)HoCo2}(Color online) Magnetic moment in
HoCo$_2$ (left) as a function of temperature obtained from the XMCD data
treatment. Open circles ({\textcolor[RGB]{2,72,142}{$\circ$}}) and open diamonds ({\textcolor[RGB]{64,120,7}{$\Diamond$}}) are the net magnetization per atom for Co and rare-earth respectively, while full diamonds {\textcolor[RGB]{253,171,2}{$\blacklozenge$}} are the sum of the
rare-earth and Co magnetization. The continuous line is the SQUID magnetization measurement. $T_c$ is the long-range magnetic ordering temperature, $T_{1}$, is the temperature at which the net magnetization of Co changes from negative to positive and $T_{2}$, is the temperature at which the net magnetization of the rare-earth sublattice changes from positive to negative}
\end{center}
\end{figure}

The temperature range in which a parimagnetic configuration takes place on DyCo$_2$, can be identified from mere inspection of Fig.~\ref{fig:XMCDDyCo2}. Indeed, it is clear that the XMCD signal at the Co L$_{2,3}$ and Dy M$_{5}$ absorption edges at $T=188$ K, (=$T_c$ + 43 K at $\mu_0H$=1 T) has the same sign as the one recorded at $T=140$ K, (5 K $<$ $T_c$ at $\mu_0H$=1 T). This indicates that on average, the Co and the Dy net moments are antiparallel within a wide range of temperature above $T_c$. As evidenced in Fig.~\ref{fig:XMCDDyCo2}, for applied field $\mu_0H$=1 T and $T>$188 K, the XMCD signal at the Co L$_{2,3}$ changes sign, remaining thereafter unchanged up to the highest studied temperature.

Above the temperature at which the net moment of Co magnetization sublattice is zero,  $T \approx 220 $ K, the disordered Co atomic magnetic moments are polarized by the applied magnetic field, creating a positive Co magnetic moment.

Analogous results are obtained in HoCo$_2$ and TmCo$_2$ (Fig.~\ref{fig:XMCDHoCo2} and Fig.~\ref{fig:XMCDTmCo2}). Well above $T_c$, at 96 K and 25 K for HoCo$_2$ and TmCo$_2$ respectively, the sign of XMCD signal for Co L$_{2,3}$ edge remains unchanged with respect to the ferrimagnetic phase, which is the telltale characteristic of parimagnetism. At a higher temperature, which depends on the magnitude of the applied field and $R$, the net Co magnetic moment changes from negative to positive. In Figs.~\ref{fig:XMCDHoCo2} and ~\ref{fig:XMCDTmCo2} it can be seen that at $T$=135 K and $\mu_0H$=2 T for HoCo$_2$ and $T$=35 K and $\mu_0H$=1 T for TmCo$_2$, the Co net magnetic moment is aligned parallel to the applied magnetic field, as expected in paramagnetism.

Up to this point, our results in $R$Co$_2$ for $R$ = Dy, Ho, and Tm show a behavior that is similar to that previously observed in ErCo$_2$\cite{Herrero-Albillos2007PRB76}. At higher temperatures, an interesting feature is found in these compounds. For example, the Dy signal in DyCo$_2$ is small, (indeed comparable to noise level) but the minute, positive signal at 1296 eV observable in the bottom right panel of Fig.~\ref{fig:XMCDDyCo2} suggests a reversal on the sign of the Dy XMCD at the M$_{5}$ edge for $T > 228$ K. The Dy sublattice magnetization would become opposite to the Co net magnetization and to the applied magnetic field. The experimental results on HoCo$_2$ and  TmCo$_2$ unambiguously confirm this trend: Ho and Tm M$_{5}$ XMCD signals change their sign at high temperature, above $T=135$ K and $T=35$ K, respectively (right bottom spectra of Fig.~\ref{fig:XMCDHoCo2} and Fig.~\ref{fig:XMCDTmCo2}). In view of these results, two temperatures at which one of the two sublattice magnetization crosses zero well above the critical temperature can be defined: $T_{1}$, where the net magnetization of Co changes from negative to positive and $T_{2}$, where the net magnetization of the rare-earth sublattice changes from positive to negative (see Table \ref{Table1:Temperatures}).

\begin{table*}[!htb]
\centering
\begin{tabular}{ccccc}
\toprule[1.3pt]
   \multirow{2}{*}{\head{Compound}} &\multirow{2}{*}{ \head{$T_c$ (K)}} &\multicolumn{2}{c}{\head{XMCD}} & \multicolumn{1}{c}{\head{TS}}  \\
    \cmidrule(lr){3-4}\cmidrule(lr){5-5}
   & & \head{$T_{1}$ (K)} & \head{$T_{2}$ (K)} & \head{$T_{1}$ (K)} \\
   \toprule[1.pt]
 DyCo$_2$ & 140 & 210$<T_1<$228 & 228$<T_2<$252 & 200 \\
  \cmidrule(lr){1-1} \cmidrule(lr){2-2} \cmidrule(lr){3-4}\cmidrule(r){5-5}
 HoCo$_2$ & 78& 88$<T_1<$112 & 135$<T_2<$200 & 104 \\
   \cmidrule(lr){1-1} \cmidrule(lr){2-2} \cmidrule(lr){3-4}\cmidrule(lr){5-5}
 TmCo$_2$ & 4.6& 25$<T_1<$32 & 38$<T_2<$41 & 20\\
\bottomrule[1.pt]
\end{tabular}
\caption{\label{Table1:Temperatures} Temperatures $T_1$ (at which the net magnetization of Co changes from negative to positive) and $T_2$ (at which the net and magnetization of the rare-earth sublattice changes from positive to negative) estimated from XMCD and $T_{1}$, obtained from TS measurements.}
\end{table*}

The separate thermal evolution of Ho and Co net magnetic moments obtained from XMCD measurements at $\mu_0H$=1 T is shown in Fig.~\ref{fig:M(T)HoCo2}. The signals in the ferrimagnetic phase have been scaled to the values of Ho and the Co moments, as reported from neutron diffraction experiments.\cite{Bloch_SSC1973}  As expected, the total magnetization measured in a SQUID magnetometer coincides with the sum of both magnetization sublattices, i.e., the sum of Ho and twice the Co net magnetization sublattices. Both Ho and Co magnetizations show abrupt jumps at the ordering transition. $T_1$ and $T_2$ are also indicated.

One would expect that under a sufficiently high applied magnetic field, the conventional paramagnetic configuration should be recovered. To complete our study of the paramagnetic phase of the $R$Co$_2$ series, we measured XMCD for applied fields from $\mu_0H$=0.25 T up to $\mu_0H$=6 T, at selected temperatures.   Fig.~\ref{fig:IsofieldHoCo2} shows the XMCD spectra recorded at the Co L$_{2,3}$ and Ho M$_{4,5}$ edges (left and right panel, respectively) measured in HoCo$_2$ at $T$=352 K. The experimental results show that a field of $\mu_0H$=4 T at room temperature is required to recover the polarization signs of conventional paramagnetism for HoCo$_2$. The Co XMCD signal is essentially unchanged, indicating that Co net magnetization is parallel to the applied magnetic field at $T$=352 K, independently of the orientations of the Ho net magnetic moment.

Four configurations, common to the three ferrimagnetic compounds for Co and $R$ net magnetic moments are identified from the XMCD measurements under applied magnetic fields up to $\mu_0H$=5 T in a temperature range between $T$=4 K to $T$=350 K : \textsl{a)} the long-range ordered antiparallel alignment below $T_c$; \textsl{b)} the parimagnetic low-field intermediate-temperature phase, denoted as \textsl{parimagnetic-1}, (which is essentially the parimagnetism observed in ErCo$_2$); \textsl{c)} the expected high-temperature paramagnetic configuration, which almost disappears at low fields, substituted by \textsl{d)} an unexpected, high-temperature parimagnetic configuration with an inverted, negative net rare-earth magnetization, denoted as \textsl{parimagnetic-2} to differentiate it from the low temperature parimagnetic one.

\begin{figure*}[!htb]
\begin{center}
\includegraphics[width=0.32\textwidth]{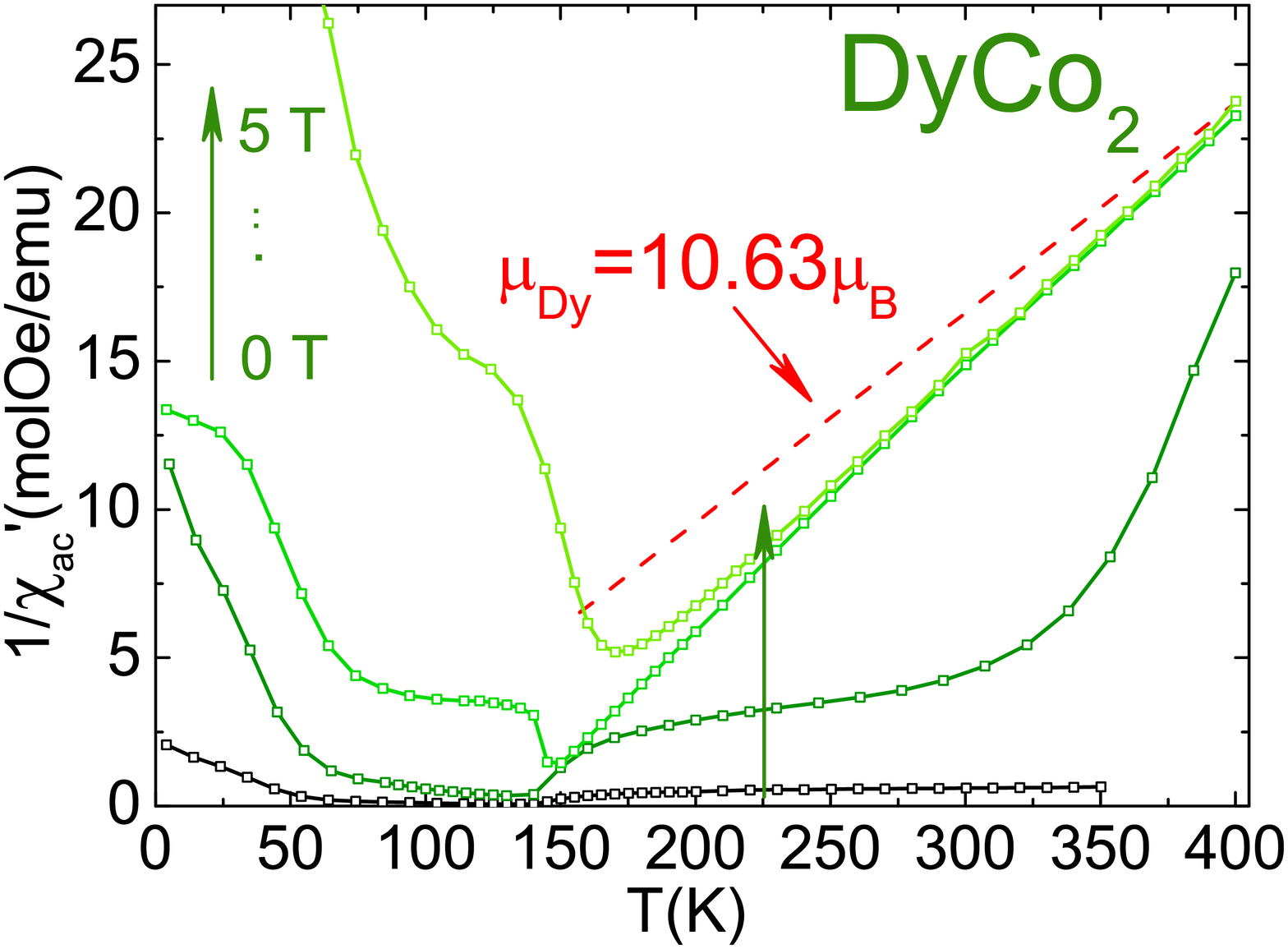}
\includegraphics[width=0.32\textwidth]{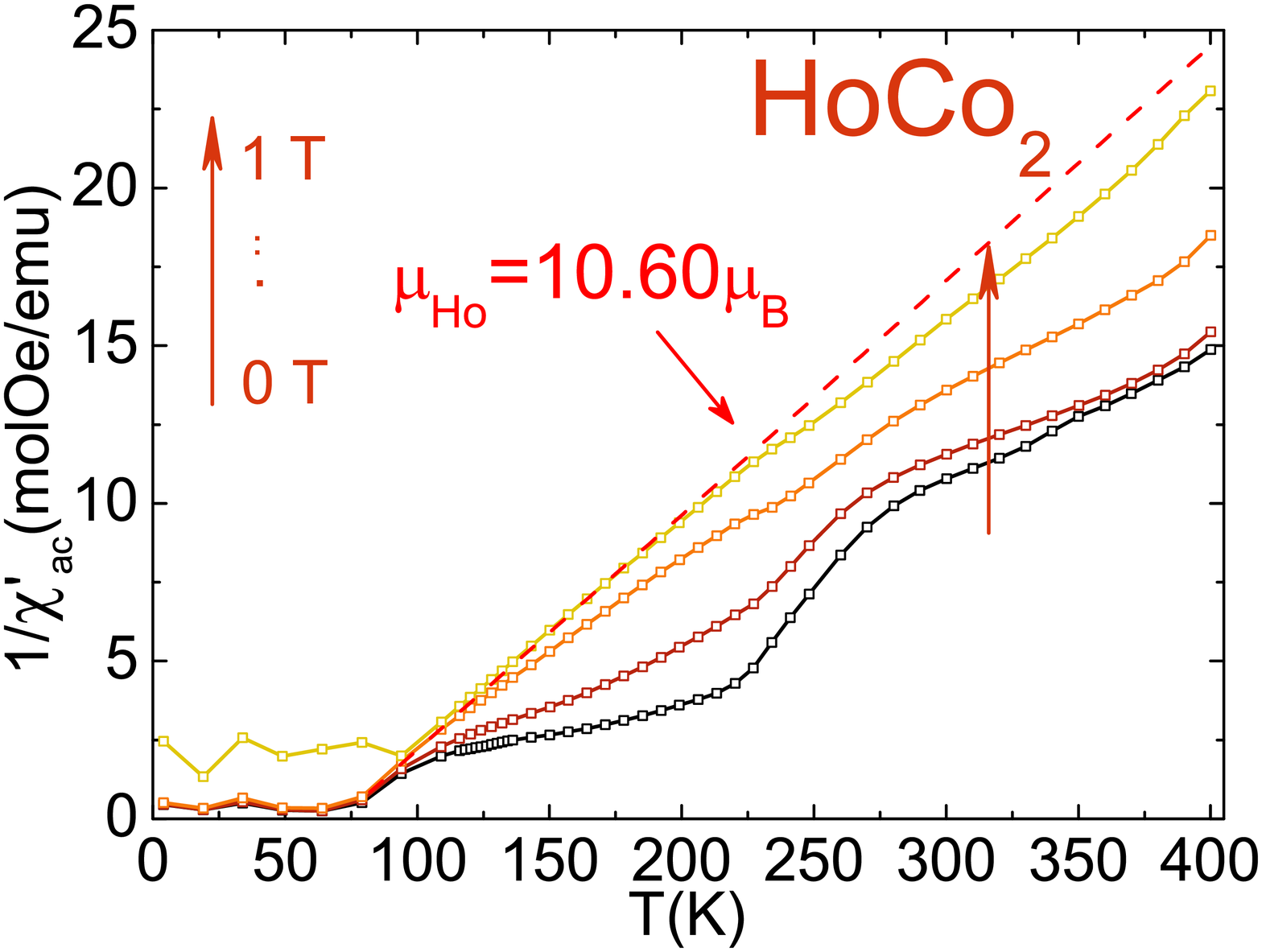}
\includegraphics[width=0.32\textwidth]{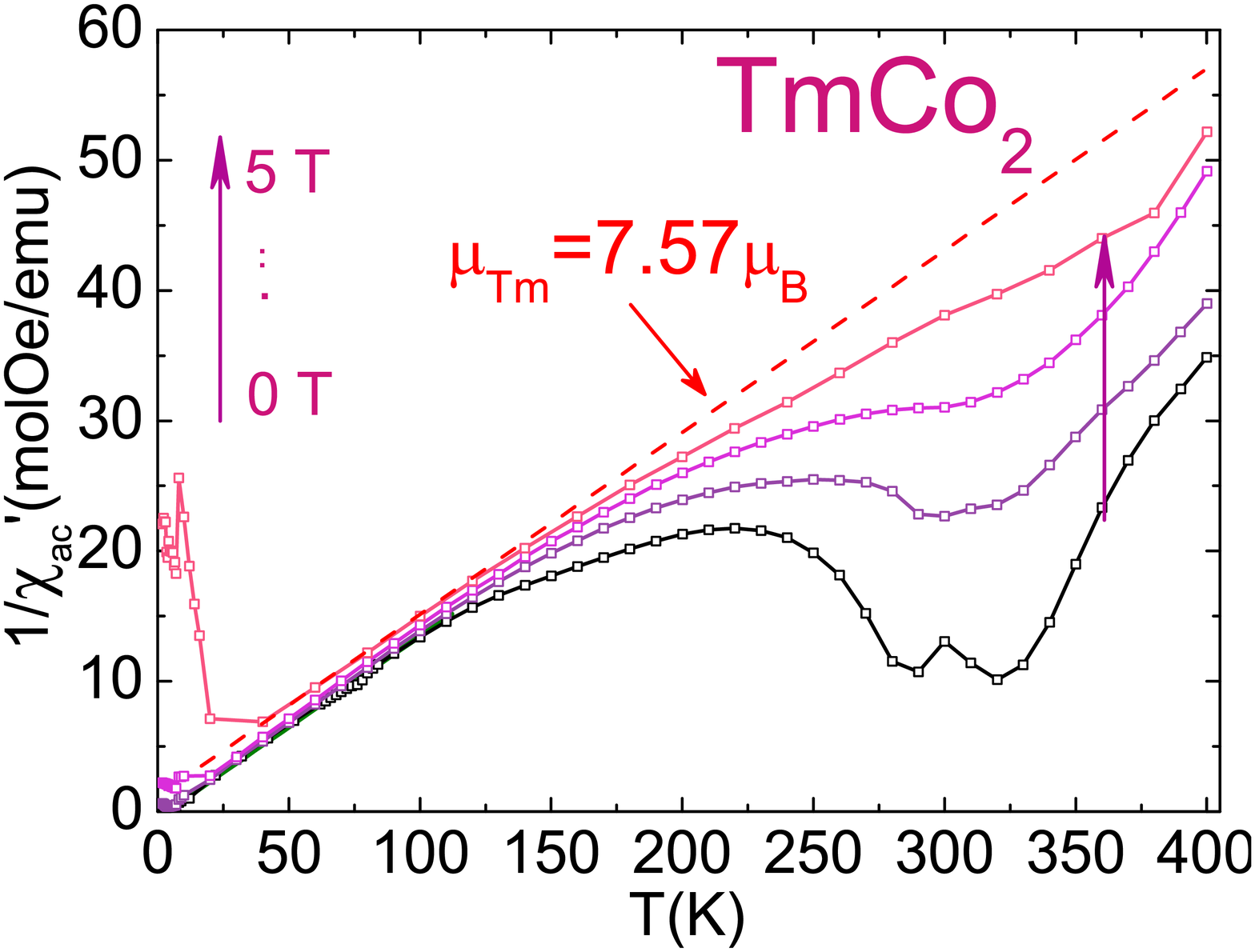}
\caption{\label{fig:XacRCo2}(Color online) Temperature dependence
of inverse of ac susceptibility. From left to right: DyCo$_2$; at $\mu_0H$=0 T, 0.04 T, 1 T and 5 T, HoCo$_2$;  at $\mu_0H$=0 T, 0.05 T, 0.3 T and 1T
TmCo$_2$;  at $\mu_0H$=0 T, 0.1 T, 1 T and 5 T. Dashed line shows the theoretical behavior expected for
each system in the paramagnetic region according to Curie-Weiss
law.}
\end{center}
\end{figure*}

\subsection{ac magnetic susceptibility}

As it has been discussed in previous works,\cite{Herrero-Albillos2007PRB76, Herrero-Albillos_Tesis, Herrero-Albillos_JPCM2009} the occurrence of parimagnetism in ErCo$_2$ is due to a competition of interactions on the Co sublattice. In particular, the Co-Co exchange interaction is responsible for the short-range order correlations which facilitate the collective reversal of the Co moments at the temperature where the net moment of the Co magnetic sublattice crosses zero. The phenomenology found by means of XMCD experiments in DyCo$_2$, HoCo$_2$, and TmCo$_2$, is similar to that observed in ErCo$_2$, and therefore short-range correlations in the paramagnetic phase of the $R$Co$_2$ systems considered in this study, are expected to occur. Indeed, previous SANS measurements have been used recently to probe the presence of short-range correlations within the magnetically disordered phase of HoCo$_2$, \cite{Bonilla_2012} yielding a correlation length of 7\,\AA~in HoCo$_2$ around $T_1$, in coincidence with our previous result in ErCo$_2$\cite{Herrero-Albillos2007PRB76}. Magnetic ac susceptibility measurements will be presented here to get insight on the origin of parimagnetism in those compounds.

The inverse of $\chi_{ac}'(T)$, for DyCo$_2$ (left), HoCo$_2$ (center) and TmCo$_2$ (right), at selected applied magnetic fields between $\mu_0H$=0 T to $\mu_0H$=5 T is shown in Fig~\ref{fig:XacRCo2}. In a standard paramagnetic system, the inverse of $\chi_{ac}'(T)$ has a linear dependence with temperature,  according to the Curie-Weiss law
\begin{equation}
\chi_{ac}'(T)=\frac{C}{T-\theta}~~~\textrm{and}~~~C=\frac{N\mu_B^2}{3k_B}\mu_{\emph{eff}}^2
\end{equation}
where N is the number of dynamic entities, $\theta$ is the Curie temperature, and $\mu_{\emph{eff}}$ is the effective magnetic moment.
$R$ magnetic moments are two orders of magnitude larger than the Co
one,\cite{GignouxPFMPhys1977,Gignoux_PRB1976} for $T>T_c$ in the $R$Co$_2$,  therefore, $\mu_{\textrm{eff}}$ is expected to be entirely due to the rare-earth moment.

Within this approximation, the dashed lines in Fig.~\ref{fig:XacRCo2} represent the contribution to the paramagnetic susceptibility from $R$ ions by using the S, L and J values predicted by the Hund's rules, and assuming that only the ground state J is populated. The estimated values for $\mu_{\emph{eff}}$ are 10.63 $\mu_B$, 10.60 $\mu_B$ and 7.57 $\mu_B$ for Dy, Ho and Tm ions, respectively. Clear deviations from those values and, in general, from a linear dependence with temperature of the $\chi_{ac}^{-1}$ can be observed for the three compounds, with the presence of one or two minima in the curves for $\mu_0H<=$1 T. Those strong deviations from the simple Curie-Weiss curve at lower applied magnetic fields corresponds to values of $\mu_{\emph{eff}}$ considerably higher than those estimated from the rare-earth magnetic moment only.  For example, the curve measured at zero applied field for HoCo$_2$ allows to extract a Curie-Weiss value of $\mu_{\emph{eff}} \sim$ 13 $\mu_B$, about 2.5 $\mu_B$ higher than the predicted value for $R$ magnetic moment only. The high-temperature XMCD signal at the Co L$_{2,3}$ edges shows a non negligible magnetic moment in Co, that is nevertheless not large enough to explain the strong $\mu_{\emph{eff}}$ observed. We ascribe the high value of $\mu_{\emph{eff}}$ to the formation of volumes of short-length correlated spins, of about 1.5 nm in diameter near $T_1$, as observed by SANS in ErCo$_2$ and HoCo$_2$\cite{Herrero-Albillos2007PRB76,Bonilla_2012}.

At high applied fields, the ac magnetic susceptibility decreases and at $\mu_0H$=5 T the expected linear trend of the inverse susceptibility for $R$Co$_2$ with small, uncorrelated Co moments is almost recovered.  This behavior, also observed in magnetically disordered ErCo$_2$, has been related with the \textsl{extreme sensitivity} to the applied field of $\chi_{ac}$ taking place at the onset of a Griffiths-like phase.\cite{Ouyang_PRB_2006,JiangPRB2007,Salamon_PRB_2003}

\subsection{Radio-frequency transverse susceptibility and magnetic phase diagrams}

\begin{figure}[!htb]
\begin{center}
\includegraphics[width=0.4\textwidth]{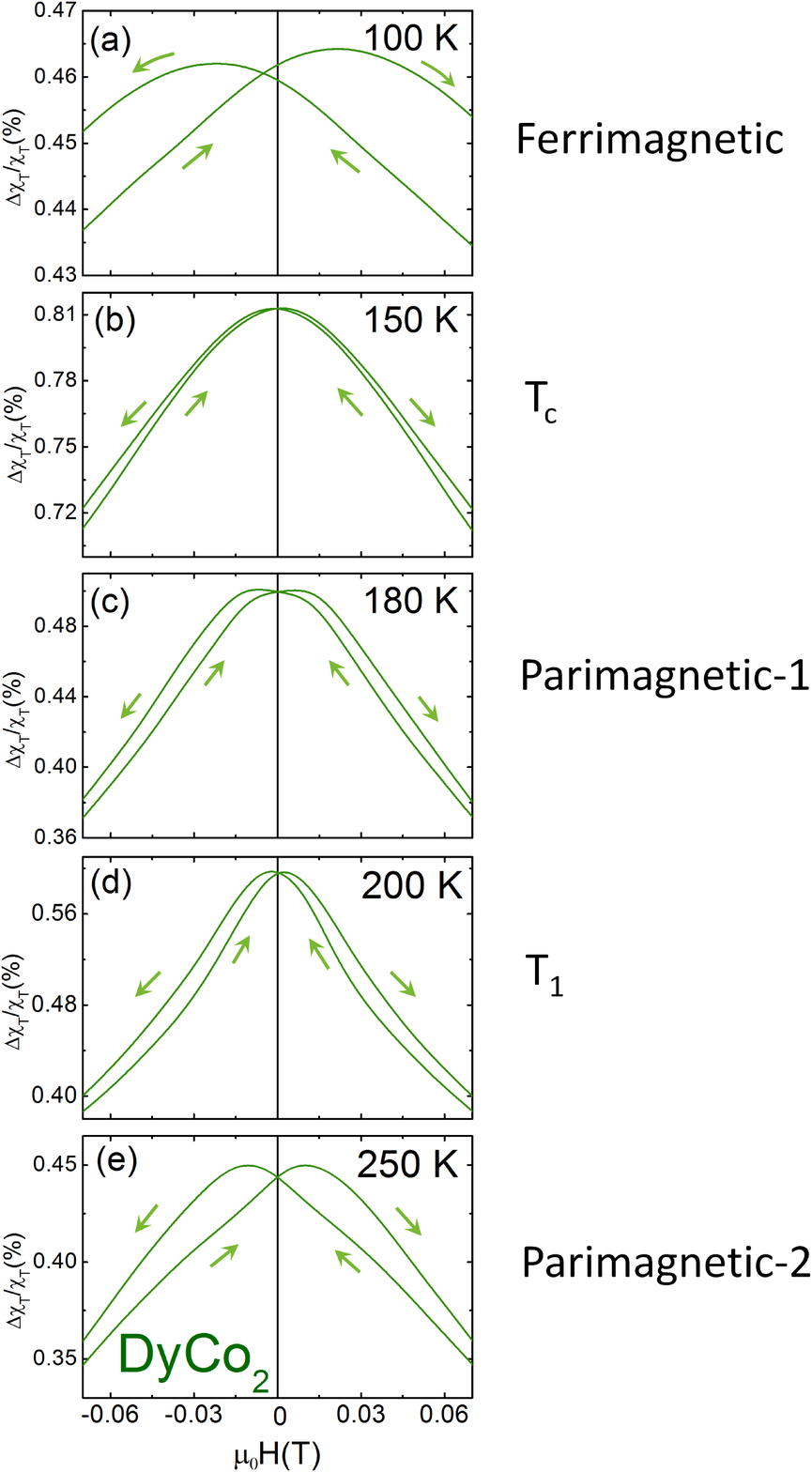}
\caption{\label{fig:TSRCo2}(Color online)  TS scans for DyCo$_2$, for $\mu_0$H$_{sat}$= 1 T  at
selected temperatures representative of the TS profile evolution in ferri- (panel a), pari-1 (c) and pari-2 (e) phases; TS bipolar scans shown in panels (b) and (d) have been recorded at $T_c$ and $T_1$, respectively.}
\end{center}
\end{figure}

Transverse Susceptibility (TS) technique has been used to study the anisotropic magnetic properties and magnetic switching in a variety of systems from multilayered thin films\cite{Frey_PRB2006} to single crystals\cite{Woods_JAP2005} and nanoparticles.\cite{Poddar_JAP2008} More interesting to our case, a careful analysis of TS profiles (their shape and magnitude) has also revealed several fascinating features, such as phase coexistence and short-range correlations that are present in doped manganites\cite{Woods_JAP2005,Chandra_JAP_2011} and cobaltites.\cite{Leighton_PRB2009,Frey_PRB2011}
Indeed, the sensitivity of this technique to short-range magnetic correlations has been previously demonstrated in $R$Co$_2$ compounds\cite{Bonilla_PRB2011,Figueroa_JAP2011}, as it provides insight of the spin dynamics in each temperature region, particularly in their magnetically disordered phase. While an unipolar TS scan recorded below $T_c$, on a polycrystalline ferromagnet has a maximum at a field, $H_M$ which coincides with the switching field, $H_S$,  which is the field needed to reverse the direction of the net magnetization; an ideal paramagnet shows $H_M$=0\cite{Chandra_JAP_2011} since the magnetic moments are disordered and uncorrelated, being able to adiabatically follow the RF excitation. In contrast, the TS bipolar scans measured above $T_c$ in ErCo$_2$ showed maxima at nonzero $H_{M}$ in the temperature range where short-range magnetic correlations are present.\cite{Figueroa_JAP2011}

In this work we present TS results performed in DyCo$_2$, HoCo$_2$ and TmCo$_2$. DyCo$_2$ bipolar scans are shown in Fig.~\ref{fig:TSRCo2} for $\mu_0$H$_{sat}$= 1 T at selected temperatures: $T$=100 K (panel a), 150 K (b), 180 K (c), 200 K (d) and 250 K (e), representative of the TS profile evolution in ferri- (panel a), pari-1 (c) and pari-2 (e) phases. TS bipolar scans shown in panels (b) and (d) have been recorded at $T_c$ and $T_1$, respectively. Panel (a) shows peaks at the switching field of ferrimagnetic DyCo$_2$. The peaks are broad, but the switching field can still be properly determined. TS bipolar scans at and above $T_c$ are narrower than that measured below $T_c$. In particular,
$H_{M}$ is almost zero at $T_c$ (panel b), but it is clearly nonzero at higher temperatures.
The finite values of $H_{M}$ found at $T>T_c$ in DyCo$_2$ suggest the occurrence of short-range correlations within the paramagnetic region. The short-range order gives rise to an increase in the field needed to reverse the magnetic moments, $H_{M}$, as they follow the RF excitation, in agreement with previous experimental works.\cite{Woods_JAP2005,Leighton_PRB2009,Figueroa_JAP2011} A description for a similar study on HoCo$_2$ can be found in Ref. \onlinecite{Bonilla_2012} where the same phenomenology has been observed up to temperatures as high as $T=300$ K.

\begin{figure}[!htb]
\begin{center}
\includegraphics[width=0.47\textwidth]{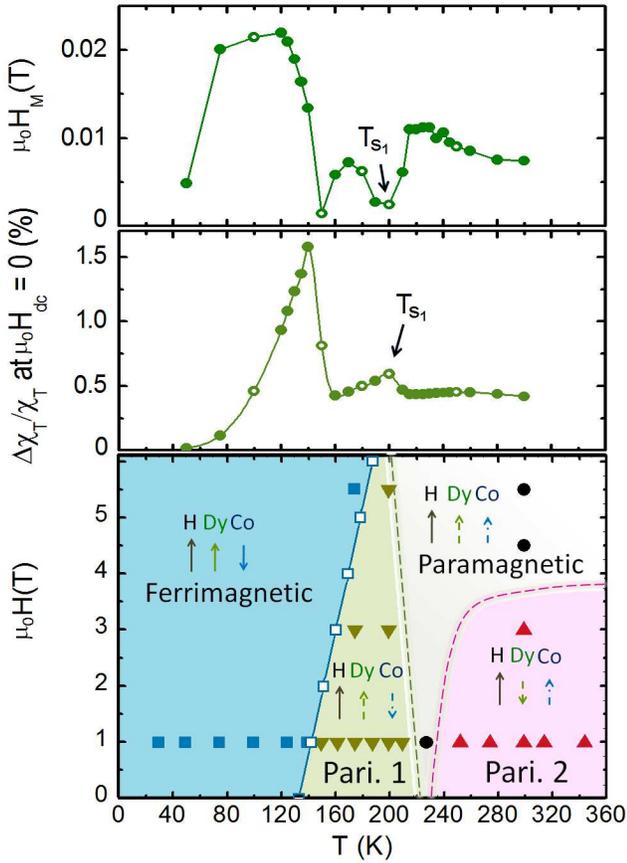}
\caption{\label{fig:SwitchingDyCo2}(Color online) Top: Field at which a maxima is found in the TS unipolar scans $\mu_0H_M$ as function of temperature for DyCo$_2$. The values of $H_M$ obtained from the selected temperatures in Fig.~\ref{fig:TSRCo2} are represented in open symbols. Middle: Susceptibility identified from TS measurements with $\mu_0H^{sat}=1$ T for DyCo$_2$ Bottom: New
magnetic phase diagram proposed for DyCo$_2$ from XMCD analysis
for different fields and temperatures. Dashed line are a guide to
the eye to separate the different configurations between the Co
and Dy net magnetic moments. {\textcolor[RGB]{0,64,128}{$\square$}} are selected values
of $T_c$ from the literature.\cite{Herrero-Albillos_PRB2006} Full symbols
indicate the field and temperature values at which XMCD
measurements have been recorded in DyCo$_2$. {\textcolor[RGB]{0,119,176}{$\blacksquare$}} are
used for the ferrimagnetic phase (Co $\downarrow$ Dy $\uparrow$).
{\textcolor[RGB]{102,130,38}{$\blacktriangledown$}}, {\textcolor[RGB]{0,0,0}{$\bullet$}} and {\textcolor[RGB]{240,15,49}{$\blacktriangle$}} are used in
the disordered region, to indicate the parimagnetic-1 (Co
$\downarrow$ Dy $\uparrow$), paramagnetic (Co $\uparrow$ Dy
$\uparrow$) and parimagnetic-2 configurations (Co $\uparrow$
Dy $\downarrow$) respectively. }
\end{center}
\end{figure}

\begin{figure}[!htb]
\begin{center}
\includegraphics[width=0.47\textwidth]{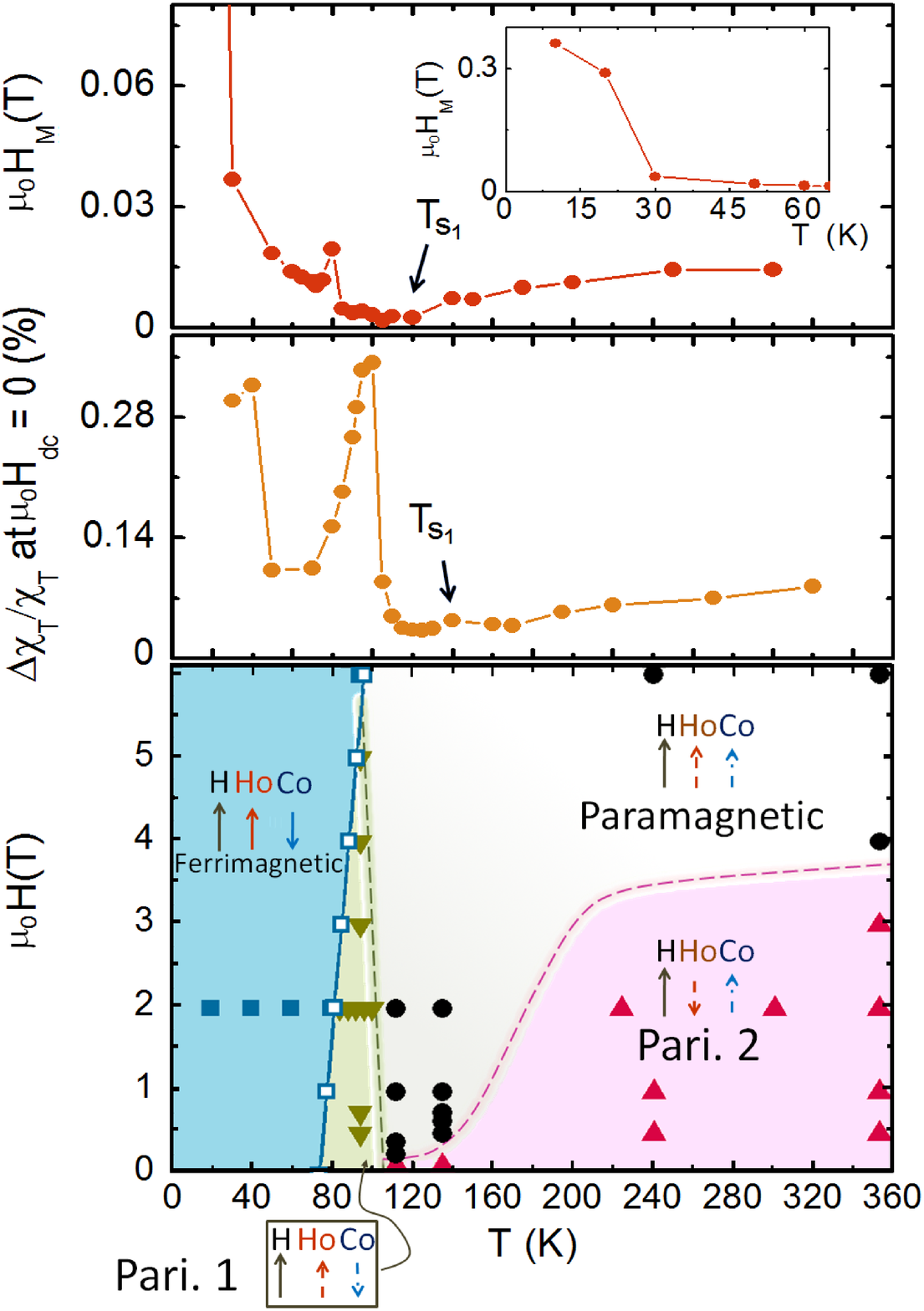}
\caption{\label{fig:SwitchingHoCo2}(Color online) Top: Field at which a maxima is found in the TS unipolar scans $\mu_0H_M$ as function of temperature for HoCo$_2$.  The inset shows the spin reorientation at low temperature in HoCo$_2$ that can be identified by a jump to very high $H_M$ values at $\sim 30$ K. Middle: Susceptibility identified from TS measurements with $\mu_0H^{sat}=1$ T for HoCo$_2$  Bottom: New
magnetic phase diagram proposed for HoCo$_2$ from XMCD analysis
for different fields and temperatures. Dashed line are a guide to
the eye to separate the different configurations between the Co
and Ho net magnetic moments. {\textcolor[RGB]{0,64,128}{$\square$}} are selected values
of $T_c$ from the literature.\cite{Herrero-Albillos_PRB2006} Full symbols
indicate the field and temperature values at which XMCD
measurements have been recorded in HoCo$_2$. {\textcolor[RGB]{0,119,176}{$\blacksquare$}} are
used for the ferrimagnetic phase (Co $\downarrow$ Ho $\uparrow$).
{\textcolor[RGB]{102,130,38}{$\blacktriangledown$}}, {\textcolor[RGB]{0,0,0}{$\bullet$}} and {\textcolor[RGB]{240,15,49}{$\blacktriangle$}} are used in
the disordered region, to indicate the parimagnetic-1 (Co
$\downarrow$ Ho $\uparrow$), conventional paramagnetic (Co
$\uparrow$ Ho $\uparrow$) and parimagnetic-2 configurations
(Co $\uparrow$ Ho $\downarrow$) respectively. }
\end{center}
\end{figure}

\begin{figure}[!htb]
\begin{center}
\includegraphics[width=0.47\textwidth]{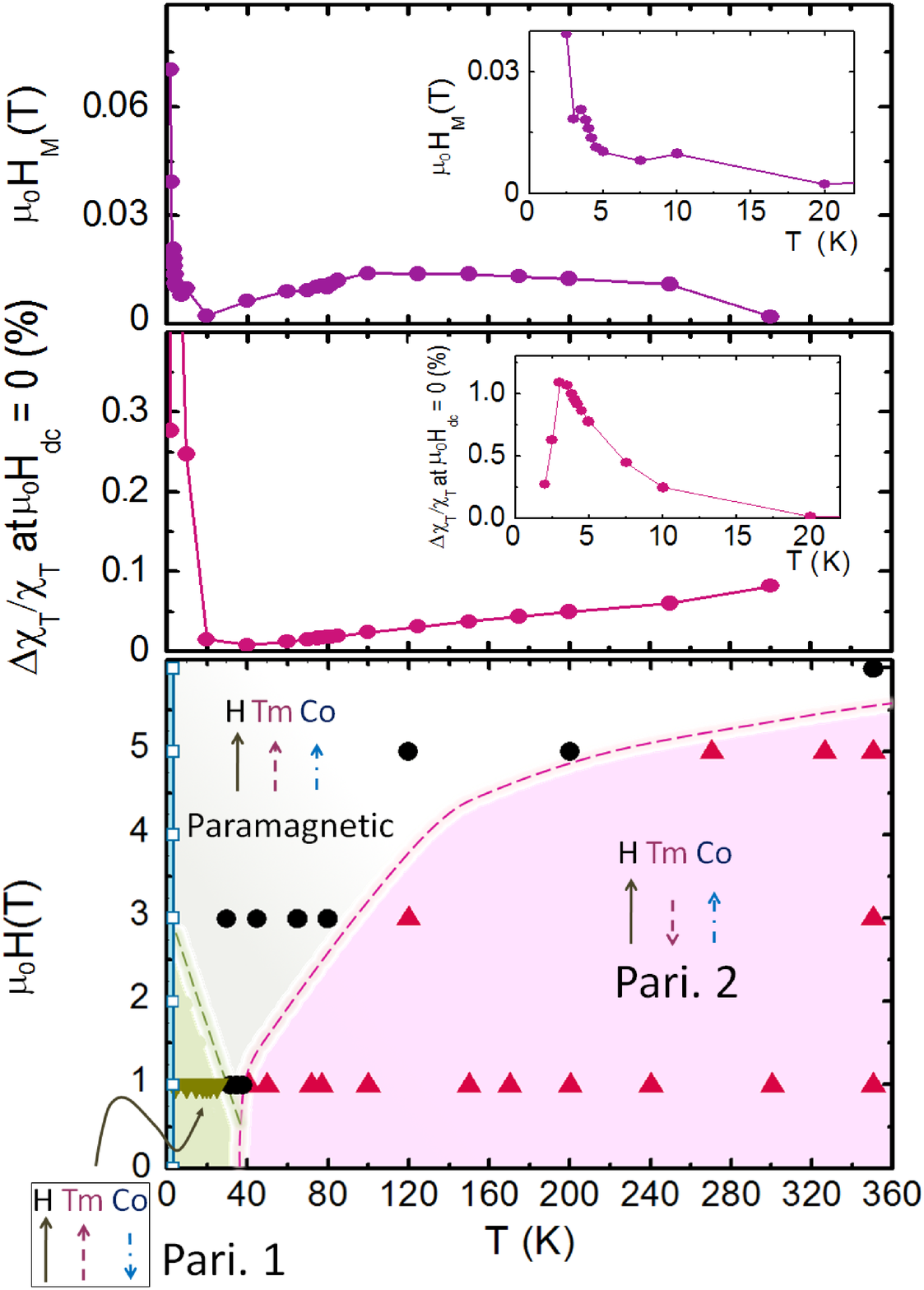}
\caption{\label{fig:SwitchingTmCo2}(Color online) Top: Field at which a maxima is found in the TS unipolar scans $\mu_0H_M$ as function of temperature for TmCo$_2$. The $H_M$ and temperature ranges of the main panel and the inset of  Fig.~\ref{fig:SwitchingTmCo2} have been chosen to allow the observation of the ordering transition at $T=4.6$ K and the evolution of $H_M$ for TmCo$_2$ above $T_c$. Middle: Susceptibility identified from TS measurements with $\mu_0H^{sat}=1$ T for TmCo$_2$.  Bottom: New
magnetic phase diagram proposed for TmCo$_2$. Dashed line are
guides to the eye to separate the different configurations between
the Co and Tm net magnetic moments. {\textcolor[RGB]{0,64,128}{$\square$}} are values
of $T_c$ from magnetization as function of temperature measurements presented in Fig.~\ref{fig:M(T)RCo2}.
{\textcolor[RGB]{102,130,38}{$\blacktriangledown$}} are the temperatures and fields at which the
alignment between Co and Tm net magnetic moment is antiparallel
for temperatures above $T_c$ (parimagnetism-1). {\textcolor[RGB]{0,0,0}{$\bullet$}} are the
temperatures and fields at which the XMCD analysis indicates a
parallel alignment between Co, Tm net magnetic moments and $H$
(conventional paramagnetism). {\textcolor[RGB]{240,15,49}{$\blacktriangle$}} are the
temperatures and fields at which the XMCD analysis indicates that
Tm net magnetic moment is antiparallel to Co net magnetic moment
and to the applied field $H$ (parimagnetism-2).}
\end{center}
\end{figure}

The temperature evolution of $H_M$ can be obtained from the analysis of the TS profiles measured at a number of temperatures, as shown in the top panel of Figs.~\ref{fig:SwitchingDyCo2},~\ref{fig:SwitchingHoCo2}, and~\ref{fig:SwitchingTmCo2} for DyCo$_2$, HoCo$_2$ and TmCo$_2$ respectively. In the three studied systems $H_{M}\neq 0$ for $T<T_c$ due to the long-range ferrimagnetic alignment between the $R$ and Co magnetization sublattices. Then, the $H_M$ profile shows a minimum at $T_c$, associated to the collective switching behavior of the rare-earth and Co moments in the ferrimagnetic ordered state. Above $T_c$, $H_{M}$ never reaches zero in the measured temperature range. This is an indication of the presence of short-range correlations in all the $R$Co$_2$ compounds, since, as it has been pointed out before, $H_M=0$ in an ideal paramagnet.

The $H_M$ curve measured for DyCo$_2$ shows two minima, one at 150 K (at $T_c$) and a second one at $T$=200 K approximately coincident with $T_1$ as obtained from the XMCD analysis. The minima in the $H_M$ profile matches with the maxima observed at the magnetic susceptibility, as it is shown in the middle panel of Fig.~\ref{fig:SwitchingDyCo2}. At temperatures above $T_{1}$, the $H_M$ curve increases slowly, and no feature is present at the temperature $T_2$ due to the inversion of the Dy moments. These observations indicate that the net magnetic moment in the field direction  due to the Co short-range correlated moments is the dominant contribution to the TS susceptibility for $T$ above $T_c$. The $R$ moments, instead, follow adiabatically the Co short-range correlated moments behavior. Magnetization measurements at temperatures above $T_c$ have been performed in DyCo$_2$\cite{Herrero-Albillos_PRB2006} corroborating that the $H_{M}$ values depicted in Fig~\ref{fig:SwitchingDyCo2} within the disordered magnetic phase do not correspond to a switching field of any spurious ferro- or ferrimagnetic phase.

The characteristic relaxation time of the fluctuation due to these short-range correlations can be calculated from Ref.~\onlinecite{Herrero-Albillos_JPCM2009} as $\tau_{SRC}\simeq 10^{-3}$ s in the temperature range of $T_1$ which is much larger than the experimental time $\tau_{exp}=1/\omega_{exp}=10^{-7}$ s. Therefore, the maxima in the RF TS at $H\neq 0$ may be observed in spite of the paramagnetism shown by DC magnetization.\cite{Herrero-Albillos_PRB2006} At $T>T_c$, the drag caused by this slow reaction to the RF excitation results in a displaced profile at each unipolar scan and the TS bipolar scans show two maxima, (at $\pm H_{M}$). As the Co net moment is reduced near the compensation temperature $T_1$ the $\tau_{SRC}$ decreases and therefore the value of $H_M$ decreases. Similar results are found for HoCo$_2$ and TmCo$_2$, but additionally, in the case of HoCo$_2$ very large values of $H_S$ can be also seen in the proximities of the spin reorientation transition that in this compound occurs at $T$=16 K.\cite{Gignoux_PRB_1975}

%

%

The correspondence between the results from the TS and XMCD measurements can be seen more clearly when comparing the top panel of Figs.~\ref{fig:SwitchingDyCo2},~\ref{fig:SwitchingHoCo2}, and~\ref{fig:SwitchingTmCo2} with their corresponding bottom panels, where new magnetic phase diagrams for DyCo$_2$, HoCo$_2$, and TmCo$_2$ are proposed from our extensive XMCD data. In the magnetic phase diagrams, schemes of the magnetic configurations of Co and $R$ moments are depicted to distinguish the ferrimagnetic, parimagnetic-1, paramagnetic and parimagnetic-2 configurations. The full symbols represent temperature and field values at which XMCD data were acquired to obtain the phase diagram. Square, inverted triangle, circle and triangle corresponds to the  ferrimagnetic, parimagnetic-1, paramagnetic and parimagnetic-2 configurations respectively, as deduced from the Co L$_{2,3}$ and R M$_{4,5}$ XMCD spectra. Open squares are selected values of $T_c$ obtained from the literature (Ref. \onlinecite{Herrero-Albillos_PRB2006}). The dashed lines are a guide to the eye that suggest a separation between the regions at which the different alignments between the Co and $R$ net magnetic moments take place on the magnetic phase diagram for $T > T_c$.

In general, the $H_M$ curve depicts the same regions observed in the phase diagrams obtained from XMCD in the three compounds. However, the temperature $T_{1}$ obtained from TS measurements, is lower than the temperature $T_{1}$ obtained from XMCD in DyCo$_2$ and TmCo$_2$. We ascribe such a discrepancy to the difference between the processes that are probed by each experiment; while XMCD is a static technique, TS is sensitive to the magnetic relaxation processes; therefore the temperature at which a given process is observed may vary. The XMCD experiments reflect $R$ and Co magnetization separately while the TS measurements senses the whole $R$Co$_2$ system. Table~\ref{Table1:Temperatures} allows to compare the compensation temperatures $T_{1}$ and $T_{2}$ obtained from XMCD at $\mu_0H$=1 T and the temperature $T_{1}$ obtained from TS experiments.

\section{DISCUSSION}

The onset of parimagnetism can be understood as a result of the internal and external magnetic fields acting on the Co sublattice. Within the mean field approximation one can consider three contributions to the total field acting on one of the Co moments:
the external applied field, $H_{\mathrm{appl}}$, the molecular field due to the $R$ sublattice, $H_{\mathrm{R-Co}}$, and the molecular field from the Co sublattice itself, $H_{\mathrm{Co-Co}}$. The direction of the applied magnetic field, $H_{\mathrm{appl}}$ defines the positive direction for magnetization. The magnitude of the molecular field for $R$ and Co sublattices depends on both the total magnetization of each sublattice and the molecular field constants: $n_{\mathrm{R-Co}}$, and $n_{\mathrm{Co-Co}}$, which are proportional to the exchange interactions, $J_{\mathrm{R-Co}}$, and $J_{\mathrm{Co-Co}}$, respectively.

The key point to understands the $R$Co$_2$ magnetism above $T_c$ lies in the fact that the Co-Co interaction ($J_{\mathrm{Co-Co}} \sim 10$ meV)\cite{RhyneJMMM1987} is much stronger than the Co-$R$ interaction ($J_{\mathrm{R-Co}}\sim$ 0.1 meV).\cite{CastetsJMMM1980} The difference of two orders of magnitude between the interactions would lead to a magnetism dominated by cobalt. However, the electronic  structure is such that the Co magnetic moment in the paramagnetic region is about two orders of magnitude smaller than the rare-earth magnetic moments \cite{Herrero-Albillos_JMMM2007}, and therefore, the interaction energies $J_{\mathrm{Co-Co}}\mu_{\mathrm{Co}}\mu_{\mathrm{Co}}$ and $J_{\mathrm{R-Co}}\mu_{\mathrm{R}}\mu_{\mathrm{Co}}$ are similar in magnitude. This equilibrium between the two internal fields may be easily destabilized by small or local variations on the crystal or electronic structure. $J_{\mathrm{Co-Co}}$ is so large that a fluctuation on the Co magnetic moment, either dynamic or static, will locally generate an exchange energy reduction by creating a short-range Co-Co ferromagnetically correlated zone. As cited before, the presence of magnetic short-range correlation have been confirmed by $\mu$SR spectroscopy in ErCo$_2$\cite{Bonilla_PRB2011} below $\sim$200 K, and  short-range correlation lengths of about 8 \AA~below $\sim$150 K have been evidenced by SANS\cite{Herrero-Albillos2007PRB76,Bonilla_2012} in ErCo$_2$ and HoCo$_2$.

Moreover, the ac magnetic susceptibility in the paramagnetic region of $R$Co$_2$ (R=Ho, Dy, or Tm) at zero field clearly does not follow the expected Curie linear dependence. The actual curve obtained by measuring a given sample is batch- and history dependent, although the general tendency in polycrystalline samples is the one shown in Fig.~\ref{fig:XacRCo2}: samples have a much larger polarizability than expected in the paramagnetic region, with susceptibilities typically two to five  times  larger than expected. The application of a (rather large) magnetic field is required to remove this extra polarizability and in some cases, $\mu_0H$=5 T are barely enough to recover the paramagnetic values. A very detailed characterization of the samples (x-ray and neutron diffraction, electron microscopy) do not show a noticeable presence of impurities which may reasonably explain the experimental behavior. However, any local deviation of stoichiometry would generate electronic and structural changes able to strongly correlate Co moments.

Indeed, density functional theory calculations to be published elsewhere\cite{Bonilla_2013} show that the formation of small Co short-range-correlations is favored near defects, vacancies, grain boundaries or electronic instabilities. This result reminds the strong spin fluctuations\cite{Nakama_PRB1999} which have been relevant to understand the paramagnetism and the transport properties of the Co Laves phases for decades. Any instability that locally enhances a Co magnetic moment generates a very strong internal $H_{\mathrm{Co-Co}}$ field, which easily may overcome the effect of the applied and $H_{\mathrm{R-Co}}$ fields, enhancing short-range correlations around it. A very interesting novelty in this work is that, in $R$Co$_2$, the Co short-range correlations not only generate parimagnetism at low field and temperature but also are able to reverse the (small) net magnetization of the rare earth sub-lattice at higher temperatures under moderate applied fields. Therefore, as long as short-range magnetic correlations are present; both configurations, parimagnetism-1 and parimagnetism-2, are explained by the competition between external and molecular fields acting on the Co moments.

As discussed in the Introduction, a Griffiths-like phase was identified in ErCo$_2$\cite{Herrero-Albillos_JPCM2009}. From the analysis of the longitudinal ac susceptibility, we find in DyCo$_2$, HoCo$_2$, and TmCo$_2$ the usual enhanced polarizability related with Griffiths phases,
while the expected value is recovered under applied magnetic field.\cite{Salamon_PRL_2002,Ouyang_PRB_2006,Magen_PRL_2006,Perez_PRB_2011} This sensitivity to the applied field is also characteristic of the formation of Griffiths phases. \cite{Salamon_PRB_2003,JiangPRB2007} The enhanced effective moment values found in the magnetically disordered phase can be ascribed to the formation of spin-correlated volumes which are in agreement with previous SANS results\cite{Herrero-Albillos2007PRB76, Bonilla_2012}, as expected in a Griffiths phase. Finally, the occurrence of parimagnetism itself requires that Co sublattice magnetization overcomes the lanthanide one, which is only possible if a relatively large number of Co moments ferromagnetically coupled form a larger, although short lived\cite{Bonilla_PRB2011, Herrero-Albillos2007PRB76}, cooperative magnetic moment. Therefore, it is plausible to propose the formation of a Griffiths-like phase above $T_c$ in the $R$Co$_2$ compounds analyzed in this paper.

However, as also mentioned in the Introduction, a Griffiths phase occurs in a magnetic diluted system and it is observed through the formation of short-range magnetic correlations in the range $T_c<T<T_{\mathrm G}$, being $T_{\mathrm G}<T_c^{'}$, the critical temperature of the undiluted system.

It is relevant to note that amorphous $R$Co$_2$ order magnetically at $T_c^{\mathrm{am}}$ well above room temperature (for example, ErCo$_2$ orders at $T_c^{\mathrm{am}} \sim 500$ K \cite{Boucher_JPFMP_1979} and $T_c^{am}$ $>$ 400 K for amorphous TbCo$_2$\cite{Hansen_1991}) Cobalt in amorphous ErCo$_2$ has a non negligible magnetic moment, and $J_{\mathrm{Co-Co}}\mu_{\mathrm{Co}}\mu_{\mathrm{Co}}$ sets the magnetic order so that $T_c^{\mathrm{am}} \approx 15\cdot T_c^{\mathrm{cryst}}$. In contrast, Co in crystalline $R$Co$_2$ has a reduced magnetic moment, as a consequence of the electronic structure in the ordered solid.\cite{Gratz_JPCM2001} One may think of the crystalline $R$Co$_2$ as a system in which magnetic Co has been ``diluted'' by the effect of the electronic structure. Therefore, $R$Co$_2$ could be considered as behaving like a magnetically ``diluted" system, where the large cobalt moments,  present in the amorphous phase, have been removed, indeed substituted by small, fluctuating magnetic moments (of the order of $0.2\mu_\mathrm{B}$ in ErCo$_2$).\cite{Herrero-Albillos_JMMM2007}
However, the Co-Co interaction is very strong ($J_{\mathrm{Co-Co}} \sim 10$ meV)\cite{RhyneJMMM1987} and it comes naturally that spin fluctuation on Co moments in $R$Co$_2$ easily induce ferromagnetically correlated volumes. As said before, Co develops a magnetic moment in $R$Co$_2$ near any imperfections or grain boundaries present in the system. These magnetic moments may add to intrinsic, dynamical spin fluctuations to generate a background density of short range order correlations. Those would be at the origin of the observed Griffiths-like phase behavior, and in some cases may be stable enough to induce parimagnetism, i.e. to reverse the net magnetic moment of the rare earth ions at temperatures above room temperature, in a magnetically disordered image of the compensated state above $T_{\mathrm{comp}}^{\mathrm{am}}$.

In summary, parimagnetism would be a consequence of a Griffiths-like phase of the undiluted, high-temperature magnetically ordered amorphous $R$Co$_2$. The new ``dilution'' mechanism at play in this Griffiths phase is not compositional, but purely magnetic, with origin in a combination of electronic-structure and local order around Co atoms. Considering the $R$Co$_2$ amorphous alloy as the ``pure'', undiluted system sets $T_c^{'}$ clearly above room temperature. Moreover, amorphous ErCo$_2$ also shows a compensation point at about $T_{\mathrm{comp}}^{\mathrm{am}} \sim 230$ K \cite{Boucher_JPFMP_1979}: above $T_{\mathrm{comp}}^{\mathrm{am}}$, Co net magnetization is larger than the Er one, whereas the opposite is true below $T_{\mathrm{comp}}^{\mathrm{am}}$. This phenomenology closely resembles what has been found in $R$Co$_2$ (for $R=$Dy, Ho, and Tm), giving raise to the parimagnetic-2 configuration.

\section{CONCLUSIONS}

Several experimental techniques have shown the existence of parimagnetism (the antiparallel configuration of the net magnetization of two
sublattices within the paramagnetic phase) among ferrimagnetic systems in the Co Laves phases family.
XMCD results in $R$Co$_2$ ($R=$ Dy, Ho, and Tm) show not only the inversion of Co magnetization sublattice at a compensation temperature $T_{1}$ well above $T_c$, as observed in ErCo$_2$. Surprisingly, the inversion of $R$ magnetization sublattice at a second compensation temperature, $T_{2}>T_{1}$, gives rise to a high temperature parimagnetic configuration, with the rare earth moment antiparallel to the Co one and to the field, which resembles the high temperature state of the magnetically ordered amorphous $R$Co$_2$.

These findings depict a new magnetic phase diagram for DyCo$_2$, HoCo$_2$ and TmCo$_2$, with two new parimagnetic configurations above $T_c$ and below an applied field threshold of about $3-5$ T. The origin of parimagnetism is understood in $R$Co$_2$ as a consequence of the relative intensity of both, the exchange interactions present in the system, and the relative magnitude of the cobalt and rare-earth magnetic moments. As the Co-Co exchange constant is two orders of magnitude stronger than the Co-$R$ one, the molecular field of Co sublattice is dominant as soon as any fluctuation locally drives the Co magnetic moment higher than its average value, given by the electronic structure: a local deviation of stoichiometry, impurities, vacancies, or even statistical spin fluctuations would generate strong correlations between Co moments, originating the formation of short-range correlated volumes. The strength of the Co-Co interaction would transmit this correlations along the system. Indeed, the detailed study of TS profiles as function of temperature has shown a finite value of $H_M$ for all temperatures and compounds, which is a clear indication of short-range correlations. On the other hand $\chi_{ac}$ measurements confirm a very high effective moment $\mu_{\mathrm{eff}}$, calculated from the $\chi_{ac}^{-1}(T)$ curves, which is congruent with the occurrence of Co short-range correlations. The recovery of the Curie behavior under applied field suggest a Griffiths-like behavior, which has also been observed in ErCo$_2$. An interpretation of parimagnetic $R$Co$_2$ as a Griffiths phase stands on an electronic structure - based dilution of the Co moment with respect to the ``high temperature phase'' represented by magnetically ordered amorphous $R$Co$_2$.

\section{ACKNOWLEDGMENTS}

The financial support of MAT2011/23791,  Aragonese IMANA (partially funded by the European Social Fund)  and the European FEDER funds is acknowledged. The authors acknowledge HZB for the allocation of synchrotron radiation beamtime. C. M. Bonilla acknowledge a Spanish MINECO grant and C. Cast\'{a}n and A. I. Figueroa acknowledge a JAE-Predoc CSIC grants.

\end{document}